# A novel variant of rhombic Penrose tiling

**Nobuhisa Fujita[1] and Komajiro Niizeki[2]**

[1] Institute of Multidisciplinary Research for Advanced Materials, Tohoku
University, Sendai 980-8577, Japan
[2] Professor Emeritus, Department of Physics, Graduate School of Science,
Tohoku University, Sendai 980-8578, Japan

E-mail: nobuhisa@tohoku.ac.jp

## Abstract

We present a novel variant of a planar quasiperiodic tiling with tenfold
symmetry, employing the same thick and thin rhombuses as the celebrated
rhombic Penrose tiling. Despite its distinct visual appearance, this new tiling
shares several key features with its predecessor, including similar vertex
environments, polygonal acceptance domains based on regular pentagons, and
an inflation/deflation symmetry associated with the golden mean as its
fundamental scaling ratio. Additional complexities arise from an increased
number of prototiles and a dual grid pattern that incorporates folded lines
alongside ordinary straight lines. This tiling exhibits a high density of a
compact decagonal motif forming a two-tiered, five-petaled flower pattern,
which spans a substantial portion of the tiling. We identify a slightly enhanced
degree of hyperuniform order compared to the standard rhombic Penrose
tiling.

Keywords: Penrose tilings, quasiperiodic tilings, rhombic tilings, decagonal symmetry, Amman bar grid

## 1 Introduction

Despite the recent discovery of *monotile tilings* based on a single prototile [1, 2], aperiodic tilings
usually employ two or more shapes as their prototiles. A well-known example is the pentagonal Penrose
tiling (P1), which uses six polygonal shapes to tile the plane: three types of regular pentagons, along with
thin rhombuses, crowns, and pentacles that fill the gaps between pentagons [3-5]. These six prototiles are
subject to specific matching constraints, often represented by bumps and dents on their edges, which enforce aperiodicity. The number of prototiles in P1 can be reduced to two by reassembling the six tile shapes — cutting them into pieces and recomposing them into new forms. Two paired sets of prototiles were thus discovered: kites and darts (P2), and thick and thin rhombuses (P3), both incorporating matching rules that enforce aperiodicity [3, 5, 6]. These three Penrose tilings are renowned for providing deep insight into the structure of quasicrystals [7-9].

In this paper, we demonstrate that the two rhombic shapes — the thick and thin rhombuses — used in the P3 tiling can also assemble into a distinct tiling with decagonal symmetry, hereafter referred to as the P4 tiling. The P4 tiling can be locally derived from the P1 tiling in a straightforward manner and shares many structural features with the P3 tiling, despite their distinct visual appearances. Along with close geometric relationships with the existing Penrose tilings, we illustrate all the fundamental properties of the P4 tiling, including the allowed vertex environments, acceptance domains, six kinds of prototiles, and the inflation/deflation symmetry under scaling by the golden mean, $\tau = (1 + \sqrt{5})/2$. Additionally, we demonstrate that each prototile of the P4 tiling can be uniquely decorated with Ammann bars, which, in a perfect P4 tiling, align with the Ammann bar grids used in the P3 tiling, highlighting the mutual local derivability of the two tilings. An assessment of the hyperuniform order reveals that the P4 tiling exhibits slightly enhanced hyperuniformity compared to the P3 tiling.

## 2    Preliminaries

*Aperiodic tilings with rhombuses*

Due to their simplicity and versatility in representing a wide range of structures, rhombuses with constant edge length are often employed as prototiles in aperiodic tilings. Many such tilings can be constructed using the dual-grid method [10, 11], originally invented by N.G. de Bruijn to produce the P3 tiling *via* pentagrids — equally spaced line grids arranged in five symmetry-related orientations [12]. This method introduces continuous degrees of freedom through parallel shifts of the line grids, resulting in a continuous family of tilings composed of the same rhombic prototiles. More specifically, previous studies have revealed contrasting properties in the two-point correlations of generalized Penrose tilings with pentagonal or decagonal symmetry obtained through this method [13-16], leading to a diverse range of physical behaviours [17]. The technique also revealed that the resulting tilings are *quasiperiodic*, as their vertices are obtained by orthogonally projecting lattice points cut out from a high-dimensional hypercubic lattice [11, 12].

Apart from the generalized Penrose tilings [13-17], the thick and thin rhombuses can form a further variety of aperiodic tilings. In the literature, quasiperiodic binary tilings with either polygonal [18-20] or fractal acceptance domains [21] have been extensively discussed. Additionally, a non-Pisot substitution rule was proposed to generate an aperiodic binary tiling lying beyond quasiperiodicity [22].

*Classification concepts*

While the full variety of aperiodic tilings is far from being uncovered [23], it is of significant interest to understand how individual tilings are related to each other. Besides point symmetry, we often rely on two key equivalence concepts to establish the relationships between aperiodic tilings or patterns. These concepts, known as *local isomorphism* and *mutual local derivability*, are briefly described as follows.

(i) *Local isomorphism (LI)*: Two tilings are locally isomorphic if and only if every finite region in one tiling is contained somewhere in the other, and *vice versa* [3, 5, 6]. In fact, the term 'P3 tiling' does not refer to a single realization of the tiling but to all members of its LI class. The same applies to other types of tilings named in the literature.

(ii) *Mutual local derivability (MLD)*: Two tilings are mutually locally derivable if and only if one tiling can be obtained from the other through local transformation, such that every element in the former is determined by consulting only a bounded region of the latter, and *vice versa* [24]. The MLD relationship between tilings naturally induces the MLD relationship between the LI classes to which they belong. All three versions of the Penrose tiling, namely, P1, P2, and P3, are equivalent in the sense of MLD.

## 3   P4 tiling

We hereby introduce the P4 tiling, which belongs to the same MLD class as the other three Penrose tilings, P1, P2 and P3. The arrangement of the rhombuses is distinct from that in the P3 tiling, as shown in Figure 1. Nevertheless, it appears as aesthetically appealing as the P3 tiling, with simple local patterns filling the entire plane. Notably, no more than two thin rhombuses can adjoin by edge sharing, which makes the distribution of thick rhombuses rather uniform. The P4 tiling is locally derived from the P1 tiling, by replacing every edge of the P1 tiling with a thin rhombus, with its acute corners coinciding with the two ends of the edge. The remaining gaps are uniquely filled with three distinct forms, consisting of one, three, and five thick rhombuses, respectively, connected by sharing edges. Conversely, the P1 tiling can be locally derived from the P4 tiling, wherein the edges of the former arise from the long diagonals of all the thin rhombuses in the latter.

Figure 2 (a) depicts the various vertex environments present in the P3 and P4 tilings, labeled according to Fig.5 of Ref. [14]. Five of these environments, labeled D, Q, K, J and S, which are 3-, 4- or 5-coordinated, are common to both tilings. However, the environments with coordination numbers of 6 and 7 are unique to each tiling: T and V are exclusive to P3, while U and W are exclusive to P4. In both tilings, the 5-coordinated environment, labeled S, appears in two distinct forms — $S_1$ and $S_2$, as shown in Figure 2(b) — which differ in the arrangements of thin rhombuses in their second coordination shells. Notably, the environment $S_1$ corresponds to the center of a decagonal motif that presents a two-tiered, five-petaled flower pattern — one that occupies a substantial portion of the entire tiling, as shown in Figure 1.

Like other quasiperiodic tilings, the P4 tiling can be described using a high-dimensional scheme. Let us first set the edge length of the rhombuses to 1, and their azimuth angles to integer multiples of $\pi/5$. Then the position $\boldsymbol{v}$ of any vertex, measured from a reference vertex at the origin, can be represented as $\boldsymbol{v} = \sum_j n_j \boldsymbol{e}_j =: [n_0 n_1 n_2 n_3 n_4]$, where $\boldsymbol{e}_j = \left(\cos(2\pi j/5), \sin(2\pi j/5)\right)$ and $n_j$ are integer indices with $j = 0$ to 4. Note that the indices $n_j$ are unique only up to the addition of $[11111]$ because $\sum_j \boldsymbol{e}_j = \boldsymbol{0}$. We shall hereafter enforce that the sum of the indices, $p = \sum_j n_j$, falls within the range $0 \le p < 5$, such that the indices will be uniquely determined. Now, let us lift each vertex $\boldsymbol{v}$ into five dimensions through $\widetilde{\boldsymbol{\varphi}} = \sum_j n_j \widetilde{\boldsymbol{\varepsilon}}_j$, where $\widetilde{\boldsymbol{\varepsilon}}_j = \left(\boldsymbol{e}_j, \boldsymbol{e}_{2j}, 1/\sqrt{2}\right)$ are five orthogonal basis vectors with equal length. Note that $\widetilde{\boldsymbol{\varphi}}$ is a lattice point of a five-dimensional hypercubic lattice (hereafter denoted $\mathbb{L}$) spanned by $\widetilde{\boldsymbol{\varepsilon}}_j$. An alternative expression $\widetilde{\boldsymbol{\varphi}} = (\boldsymbol{v}, \boldsymbol{v}', p/\sqrt{2})$ with $\boldsymbol{v}' = \sum_j n_j \boldsymbol{e}_{2j}$ asserts that $\widetilde{\boldsymbol{\varphi}}$ ($\in \mathbb{L}$) belongs to a hyperplanes of $\mathbb{L}$ at level $p$, wherein $\boldsymbol{v}'$ represents the complementary coordinates of $\widetilde{\boldsymbol{\varphi}}$ normal to the tiling plane.

The entire set of vertices in a rhombus tiling is a disjoint union of five subsets, denoted $\{\boldsymbol{v}\}_p$, each corresponding to a vertex class $p$ ($\in \{0, 1, 2, 3, 4\}$). $\{\boldsymbol{v}\}_p$ can be lifted up to a set of lattice points in the $p$-th lattice hyperplane of $\mathbb{L}$, where each lattice point has complementary components $\boldsymbol{v}'$ lying within the so-called acceptance domain (AD) — a bounded region in the complementary plane. The ADs for the Penorse P3 tiling are illustrated in Figure 3(a). The AD $V_1$ (or $V_4$) for $p = 1$ (or $p = 4$) is a regular pentagon of unit radius, with vectors $\boldsymbol{e}_{2j}$ (or $-\boldsymbol{e}_{2j}$) pointing towards its five vertices from its center. The AD $V_2$ (or $V_3$) for $p = 2$ (or $p = 3$) is geometrically similar to $V_4$ (or $V_1$), but scaled by a factor of $\tau$. The AD $V_0$ for $p = 0$ is empty. Each AD is dissected into sub-domains corresponding to the eight distinct vertex environments, as illustrated in the figure [12, 14].

As for the P4 tiling, there are five polygonal ADs, denoted $W_p$ for $p = 0$, 1, 2, 3 and 4, as shown in Figure 3(b). $W_0$ is a regular decagon with a radius of $2\sin(\pi/5)$, which is equal to the edge length of the pentagonal AD $V_1$ (or $V_4$) for the P3 tiling. It is subdivided into regions corresponding to environments J (isosceles triangles), U (quadrilaterals) and W (quadrilaterals), all of which involve acute corners of thin rhombuses. Thus, $W_0$ corresponds to the vertices of the parent P1 tiling illustrated in Figure 1. The AD $W_1$ (or $W_4$) for $p = 1$ (or 4) is a windmill-shaped polygon having five fan-like extensions, where the vectors from the center towards its five concave corners are $\boldsymbol{e}_{2j}$ (or $-\boldsymbol{e}_{2j}$). In $W_1$ (or $W_4$), five pentagonal sections for environment D encircle a central pentacle, which is subdivided into sections for environment Q (quadrilaterals), K (isosceles triangles) and $S_2$ (a central pentagon). Note that environments D, Q, and K are those that involve obtuse corners of thin rhombuses. Last but not least, the AD $W_2$ (or $W_3$) for $p = 2$ (or 3), which corresponds exclusively to environment $S_1$, is a regular pentagon of radius $1/\tau$, with vectors $\boldsymbol{e}_{2j}/\tau$ (or $-\boldsymbol{e}_{2j}/\tau$) pointing towards its five vertices from its center.

It is important to note that the area of a region within the ADs is strictly proportional to the density of vertices it generates. Since the total areas of the ADs are $5 \sin{(2\pi/5)}(1 + \tau^2)$ for both the P3 an P4 tilings, the vertex densities — as well as those of the rhombuses — are the same in both tilings. However, type-$S_1$ vertices alone occur $\tau^2$ times more frequently in the P4 tiling than in the P3 tiling, as the ADs $W_2$ and $W_3$ are $\tau$ times larger than the centermost sections for environment $S_1$ in $V_4$ and $V_1$, respectively. The high density of local 5-fold symmetry centers likely contributes to the aesthetic appeal of the P4 tiling.

## 4   MLD relationship between the P3 and P4 tilings

Both the P3 and P4 tilings employ thick and thin rhombuses as prototiles, differing only in their arrangement. Since P1 is MLD with both P3 [3, 5] and P4 (as shown in Figure 1), it follows that P3 and P4 are MLD with each other. Here, we illustrate how the P3 tiling can be locally derived from the P4 tiling, and vice versa.

Notice that all type-$S_1$ vertices in the P4 tiling are located at the pentagonal centres in the P1 tiling shown in  Figure 1. These vertices correspond to the two pentagonal ADs, $W_2$ and $W_3$. By connecting these vertices with edges of length $\tau^2$, one obtains a tiling composed of hexagon-, boat- and star-shaped polygons. These polygons, as illustrated in Figure 4(a), can be decomposed into rhombuses by placing a single vertex in the interior. The P3 tiling (with edge length $\tau^2$) will result from this process, provided the interior vertex is chosen appropriately.

Indeed, the boat (and respectively, star) polygon is uniquely decomposed, with the interior vertex placed at the vertices of type K (and respectively, $S_2$) in the original P4 tiling. On the other hand, the hexagon can be decomposed by placing a vertex at either of the two type-Q vertices within its interior. The right choice is such that the two thin rhombuses generated will cap the concave angles of adjacent boat and/or star tiles. This would also mean that the interior vertex within the hexagon should accompany vertices of type K and/or S$_2$ at a distance of $\tau^3$; otherwise, vertex environments not accepted in the P3 tiling will appear.

The ADs for the interior vertices within the hexagon, boat and star polygons thus determined are small pentagonal sub-domains of $W_1$ and $W_4$, as indicated in Figure 5(a) for $W_1$ by pink dashed lines. Note that these ADs, denoted hereafter $P_1$ and $P_4$, respectively, inscribing the pentacles composed of the sections for the S$_2$ and K vertices as shown in Figure 3(b), are $1/\tau$ times as small as $W_3$ and $W_2$, respectively. Altogether, $P_1(= V_4/\tau^2)$, $W_2(= V_3/\tau^2)$, $W_3(= V_2/\tau^2)$ and $P_4(= V_1/\tau^2)$, form the four ADs accepting the vertices of the $\tau^2$-scaled P3 tiling. Conversely, the P4 tiling is locally derivable from the $\tau^2$-scaled P3 tiling if the rhombic prototiles of the latter are decomposed in the same manner as revealed in Figure 4(a).

It is worth noting that another, $\tau$-scaled P3 tiling can be derived by decomposing the $\tau^2$-scaled P3 rhombuses using the well-known substitution rules for the P3 tiling. Accordingly, in Figure 4(b), the pink rhombuses are decomposed into blue rhombuses of size $\tau$. Here, each of the newly generated vertices incorporates an obtuse corner of a thin rhombus in the P4 tiling. The corresponding ADs for $p = 1$ and 4 are thus larger pentagonal sub-domains $B_1$ and $B_4$ of $W_1$ and $W_4$, respectively, painted blue in Figure 5(a). It follows that the ADs for the $\tau$-scaled P3 tiling are given by $B_1(= -V_2/\tau)$, $W_2(= -V_4/\tau)$, $W_3(= -V_1/\tau)$ and $B_4(= -V_3/\tau)$. Conversely, the rhombuses in the $\tau$-scaled P3 tiling are uniquely decorated by the vertices of the P4 tiling, allowing the latter to be locally derived from the former. However, the rhombuses are decomposed in several different manners.

It is important to note that $B_1$ and $B_4$ encompass the entire sections for vertex types Q, K and S$_2$ in the P4 tiling, but the type-D vertices are incorporated only in part. Indeed, one can confirm in Figure 4(b) that a vertex of type D is accepted in $B_1$ or $B_4$ if and only if it is not connected to another vertex of type D by two consecutive edges in the same direction. In Figure 5(a) the complementary sub-domain $Y_1$ to $B_1$ in $W_1$ is painted yellow, while $W_2$ is painted orange. The corresponding sub-domains in $W_4$ as well as $W_3$ are colored likewise. The colors defined for the vertices with $p = 1, 2, 3$, and 4 help us identify distinct kinds of rhombuses that are uniquely transformed through the $\tau$-scale inflation process, as described in the next section.

## 5   Inflation/deflation symmetry under $\tau$-scaling

For the $\tau$-scale inflation process, the P4 tiling is first expanded by the ratio of $\tau$, the golden mean. Next, the expanded tiles are decomposed into original tiles (with edge length 1; see Supplementary Figure 1), resulting in a new rhombus tiling that is locally isomorphic to the original P4 tiling. Importantly, the transformation is not unique to the shapes of the rhombuses, in stark contrast to the P3 tiling, where thick (or thin) rhombuses are transformed in the same manner.

From Figure 5(b), one can observe that thick rhombuses in the P4 tiling can be classified into three types based on their colored vertices. We will distinguish these three types by labeling them as *a*, *b*, and *c*. Specifically, a thick rhombus labeled '*a*' has two yellow vertices at its obtuse corners and one orange vertex at one of the acute corners. A rhombus labeled '*b*' has one yellow vertex and one blue vertex at its obtuse corners, along with one orange vertex at one of the acute corners. If neither of the obtuse corners are colored while both vertices at the acute corners are colored blue, it is labeled '*c*'. As shown in Figure 6 (top row), these three kinds of thick rhombuses are transformed uniquely through the $\tau$-scale inflation process.

On the other hand, all thin rhombuses have their vertices colored in the same way, with one yellow vertex and one blue vertex at their obtuse corners. In the inflation process, each thin rhombus is transformed into a complex of three thick rhombuses; however, the types of the resulting thick rhombuses are not uniquely determined. In fact, thin rhombuses can be classified according to the types of the two thick rhombuses that share the yellow vertex. There are thus three types of thin rhombuses, labeled *d*, *e*, and *f*, each of which is transformed uniquely under inflation, as shown in Figure 6 (bottom row).

It is worth noting that vertex colors are not preserved under inflation, but transform according to the following rules:

1. Uncolored vertices remain uncolored after the process.

2. Orange vertices become blue, and blue vertices become orange.

3. Yellow vertices cease to be vertices, as they correspond to interior points within thick rhombuses of type '*a*'.

These rules can be fully explained in terms of the hyper-scaling transformation induced by the $5 \times 5$ scaling matrix $\boldsymbol{S} = (S_{ij})$, defined as:

$$\tau \boldsymbol{e}_i = -(\boldsymbol{e}_{i-2 \bmod 5} + \boldsymbol{e}_{i+2 \bmod 5}) = \sum_j S_{ij} \boldsymbol{e}_j.$$

The hyper-scaling transformation $\tilde{\tau}$ defined through $\tilde{\tau}(\tilde{\boldsymbol{\varepsilon}}_i) \coloneqq \sum_j S_{ij} \tilde{\boldsymbol{\varepsilon}}_j$ transforms an arbitrary lattice point $\tilde{\boldsymbol{\varphi}} = (\boldsymbol{v}, \boldsymbol{v}', p/\sqrt{2})$ in $\mathbb{L}$ into $\tilde{\tau}(\tilde{\boldsymbol{\varphi}}) = (\tau \boldsymbol{v}, -\boldsymbol{v}'/\tau, -2p \ (mod \ 5)/\sqrt{2})$. It follows that the AD $W_0$ for uncolored vertices will be transformed into $-W_0/\tau \equiv W_0/\tau \ (\subset W_0)$, so that the resulting vertices are also uncolored. The ADs $O_2 \ (= W_2)$ and $O_3 \ (= W_3)$ for orange vertices will be transformed into $-O_2/\tau = P_1$ $(\subset B_1)$ and $-O_3/\tau = P_4 \ (\subset B_4)$, respectively, so that the resulting vertices are blue colored. The ADs $B_1$ and $B_4$ for blue vertices will be transformed into $-B_1/\tau = O_3$ and $-B_4/\tau = O_2$, respectively, so that the resulting vertices are orange colored. And finally, the ADs $Y_1$ and $Y_4$ for yellow vertices will be transformed into $-Y_1/\tau \ (\subset \overline{O_3})$ and $-Y_4/\tau \ (\subset \overline{O_2})$, respectively, falling outside the ADs for $p = 2$ and 3.

Each prototile, when scaled by a factor of $\tau$, is decomposed into a certain number or fraction of original-sized prototiles, as illustrated in Figure 6. This process transforms the counts of the six prototiles — denoted $n_a, n_b, n_c, n_d, n_e$, and $n_f$ — as $(n_a, n_b, n_c, n_d, n_e, n_f) \rightarrow (n_a, n_b, n_c, n_d, n_e, n_f).M^t$, where

$$M = \begin{pmatrix} \dfrac{1}{\tau^2} & \dfrac{1}{2\tau^2} & 1 & \dfrac{1}{\tau} & \dfrac{\sqrt{5}}{2} & \tau \\ 0 & \dfrac{3}{2} & 1 & 1 & \dfrac{1}{2} & 0 \\ 1 & 0 & 0 & 0 & 0 & 0 \\ 0 & 1 & 1 & 0 & 0 & 0 \\ 2 & 0 & 0 & 0 & 0 & 0 \\ 0 & \dfrac{1}{2} & 0 & 0 & 0 & 0 \end{pmatrix}$$

is the $6 \times 6$ inflation matrix characterizing the scaling behavior of the P4 tiling. Note that the leading eigenvalue of $M$ is $\tau^2$, reflecting the fact that both the number and area of the rhombuses scale by $\tau^2$ with each round of the inflation. The relative frequencies of the six prototiles in the P4 tiling correspond to the right eigenvector of $M$ associated with the leading eigenvalue and are proportional to $(\tau^3, 2\tau^2, \tau, \tau^2, 2\tau, 1)$.

# 6    Dual grid of the P4 tiling

Each of the two prototiles of P3 can be uniquely decorated with straight line segments, known as Ammann bars [25], named after their inventor, Robert Ammann [25]. When the matching rules are fully respected throughout the tiling, these bars align seamlessly across tile edges, forming continuous straight lines oriented along the five edge directions of a regular pentagon. Collectively, these lines form what is known as the *Ammann bar grid*. In this grid, each set of parallel lines forms a Fibonacci quasilattice, with long ($\ell_1$) and short ($\ell_2 = \ell_1/\tau$) intervals between adjacent lines. In Figure 7(a), the Ammann bar grid is shown superposed on the P3 tiling [26-28], which is assumed to have an edge length of $\tau$. The same Ammann bar grid is reproduced in Figure 7(b), this time superposed on the corresponding P4 tiling, which has an edge length of 1 (see Figure 4(b) and Figure 5(b)). Observe that each of the six prototile rhombuses is associated with a unique pattern of line segments, which continues seamlessly into similar patterns on adjacent rhombuses, thereby defining matching rules of the P4 tiling.

It is well established that analogous Ammann bar grids with non-crystallographic point symmetries can serve as an underlying framework for producing quasiperiodic tilings, which are the geometric duals of these grids. A number of examples — particularly rhombic quasiperiodic tilings with eight-, ten-, and twelve-fold rotational symmetries — have been presented in association with Ammann bar grids [29-33]. As a prototypical example, the Ammann bar grid presented above can be dualized to generate the P3 tiling

— or more precisely, a once-deflated version with unit edge length, relative to the version shown in Figure 7, which has an edge length of $\tau$.

Although the P4 tiling is not the dual of any known Ammann bar grid, it can nonetheless be interpreted as the dual of a certain grid pattern. To illustrate how this dual grid can be constructed, we begin by drawing a straight line through the center of each thin rhombus in the P4 tiling, oriented perpendicular to one of its two sets of parallel edges. In Figure 8(a), vertical blue lines pass through the centers of the thin rhombuses with horizontal edges, which are shaded in grey. Extending this process across the entire tiling generates an infinite set of parallel lines, spaced at three intervals — $\ell_1$, $\ell_2$, and $\ell_3 (= \ell_1 - \ell_2)$ — such that $\ell_1/\ell_2 = \ell_2/\ell_3 = \tau$, thereby defining a ternary quasilattice. Furthermore, repeating this construction in all five edge-normal directions produces the grid pattern shown in Figure 8(b), which turns out to be closely related to the original Ammann bar grid.

In fact, the new grid lines, shown in blue, are precisely aligned with the original Ammann lines, depicted as dashed purple lines in Figure 8(a). Specifically, each large interval $\ell_1$ between adjacent Ammann lines is bisected by a single blue line, producing a corresponding array of rhombuses aligned parallel to it. In contrast, each short interval $\ell_2$ contains a pair of blue lines — separated by a distance $\ell_3$ — centered within the interval. Notably, each such $\ell_3$-pair generates only one array of rhombuses, arising from *active intersections* on either of the two grid lines, while disregarding extraneous intersections that do not correspond to any rhombuses in the actual tiling. Each line in an $\ell_3$-pair is divided into active and inactive segments, with the active segments on the two lines being mutually exclusive. Furthermore, when two $\ell_3$-pairs intersect at a 72° angle, the active side alternates: an active segment terminates at the obtuse corner of a rhombus formed by the intersection, and a new active segment begins at the opposite corner. Figure 8(b) shows that a thick rhombus of type $c$ emerges at each such intersection. In Figure 8(c), the active segments of the $\ell_3$-pairs are shown as red lines, while dashed lines indicate the inactive segments. The active segments form a pattern of folded lines which, together with the remaining blue lines, constitute the dual grid of the P4 tiling.

## 7   Discussion

The P4 tiling may offer an advantage over the P3 tiling due to its higher density of type-$S_1$ vertices, which correspond to the centers of the two-tiered, five-petaled decagonal motifs shown in Figure 2(b), left, thereby contributing to its overall visual uniformity. It is worth noting that the only regions not covered by these decagons are exclusively occupied by thick rhombuses of type $c$. Consequently, the total areal coverage of the plane by the decagons amounts to $2/\sqrt{5} \approx 89.4427\%$.

The visual properties of the P4 tiling can be further elaborated by characterizing it in terms of *hyperuniformity* — a notion describing spatial patterns that exhibit vanishing density fluctuations at infinite wavelength [34]. A point pattern in $d$-dimensional space is said to be hyperuniform — or to exhibit hyperuniform order — if its local density fluctuations, quantified by the number variance $\sigma^2$ of points within a disc-shaped observation window of radius $R$, grow more slowly than the mean number of points in the window as $R$ increases; that is, $\sigma^2$ asymptotically behaves as $\sigma^2 \propto R^\alpha$ with $\alpha < d$. More specifically, the pattern is said to be *strongly hyperuniform* if $\sigma^2 \cong BR^{d-1}$. It has been reported that the vertex sets of quasiperiodic tilings are often strongly hyperuniform [15, 16, 35-38]. In such cases, different quasiperiodic tilings can be contrasted by the asymptotic coefficient $B$, which serves as a quantitative metric for the degree of hyperuniform order [34].

Let us compare the P3 and P4 tilings in terms of the hyperuniform order metric $B$ for their vertex sets. A large disc-shaped patch of each tiling was generated, and the number variance $\sigma^2(R)$ was numerically evaluated using observation windows centered at 62,500 randomly sampled points within a disc of radius 250 (in units of the rhombic edge length) located at the center of the patch. The order metric $B$ for each tiling was estimated by fitting the data for $\Lambda(R) \coloneqq \sigma^2(R)/R$ within $10 < R < 240$ to the asymptotic functional form $\Lambda(R) \approx B + C/R$ using the linear least squares method (see Supplementary Figure 2 for further details of the analysis). The normalized order metric $B/\sqrt{\phi}$, where $\phi = \pi\rho/4$ and $\rho$ is the number density of points, was found to be 0.5914 and 0.5903 for the P3 and P4 tilings, respectively. The result for the P3 tiling agrees with previous estimates [15, 16, 35, 38]. These values also indicate that the hyperuniform order of the P4 tiling is marginally higher — by approximately 0.2% — than that of the P3 tiling. Notably, the P3 tiling was previously reported to possess the highest hyperuniform order among the generalized rhombic Penrose tilings [15, 16].

In summary, we have illustrated the fundamental properties of the P4 tiling, introduced as a novel variant of the Penrose tilings constructed from thick and thin rhombuses. These properties include its local vertex environments, acceptance domains, MLD relationships with standard Penrose tilings, inflation rules, and the associated dual grid. Notably, the P4 tiling resembles its predecessor, the P3 tiling, in several key aspects — such as most of the allowed vertex environments, polygonal acceptance domains based on regular pentagons, and the characteristic $\tau$-scaling property — while introducing additional complexities through an increased number of prototiles and a dual grid incorporating folded lines alongside ordinary straight lines. The visual uniformity and overall structural coherence of the P4 tiling have been examined in terms of the high density of compact decagonal motifs exhibiting local fivefold symmetry and a numerical evaluation of the hyperuniform order metric. We have shown that the P4 tiling exhibits slightly higher hyperuniform order than the P3 tiling.

## Acknowledgements

The authors are grateful to Akihisa Koga for commenting on the computation of hyperuniformity order metric for quasiperiodic patterns. This work was supported by Japan Society for the Promotion of Science through Grant-in-Aid for Scientific Research (Grant No. JP19H05819) and Japan Science and Technology Agency through CREST (Grant No. JPMJCR22O3).

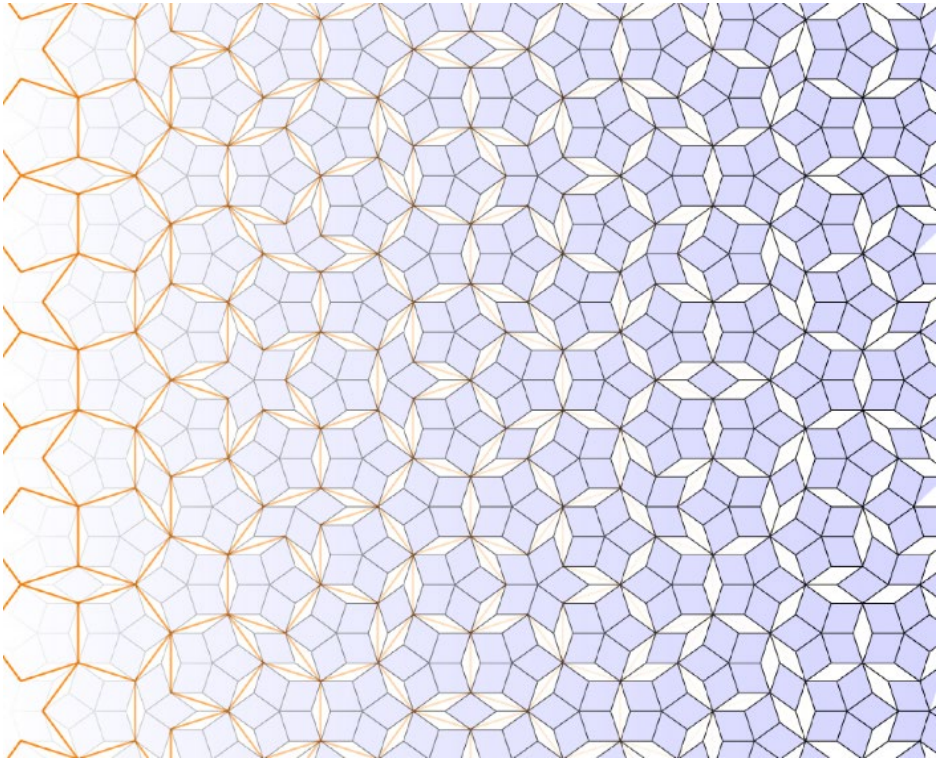

***Figure 1:*** **The P4 tiling composed of thick (shaded) and thin (white) rhombuses. On the left, the derivation of this tiling from the P1 tiling (drawn using orange solid lines) is shown.**

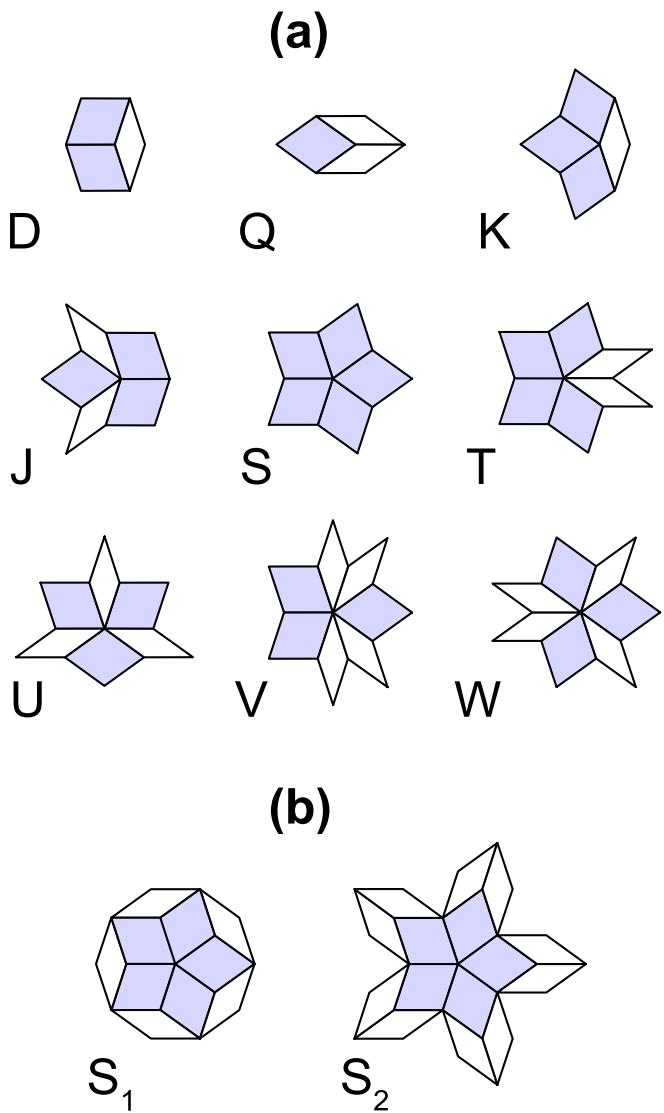

**(a)**

D    Q    K

J    S    T

U    V    W

**(b)**

$S_1$    $S_2$

*Figure 2:* (a) The vertex environments allowed in the P3 and P4 tilings. Whereas five low-coordinated environments — D, Q, K, J and S — are common in both tilings, hexa- or hepta-coordinated environments are exclusive to one another: T and V in P3, and U and W in P4. (b) Two kinds of coordination clusters up to the second shell of the penta-coordinated environment, S.

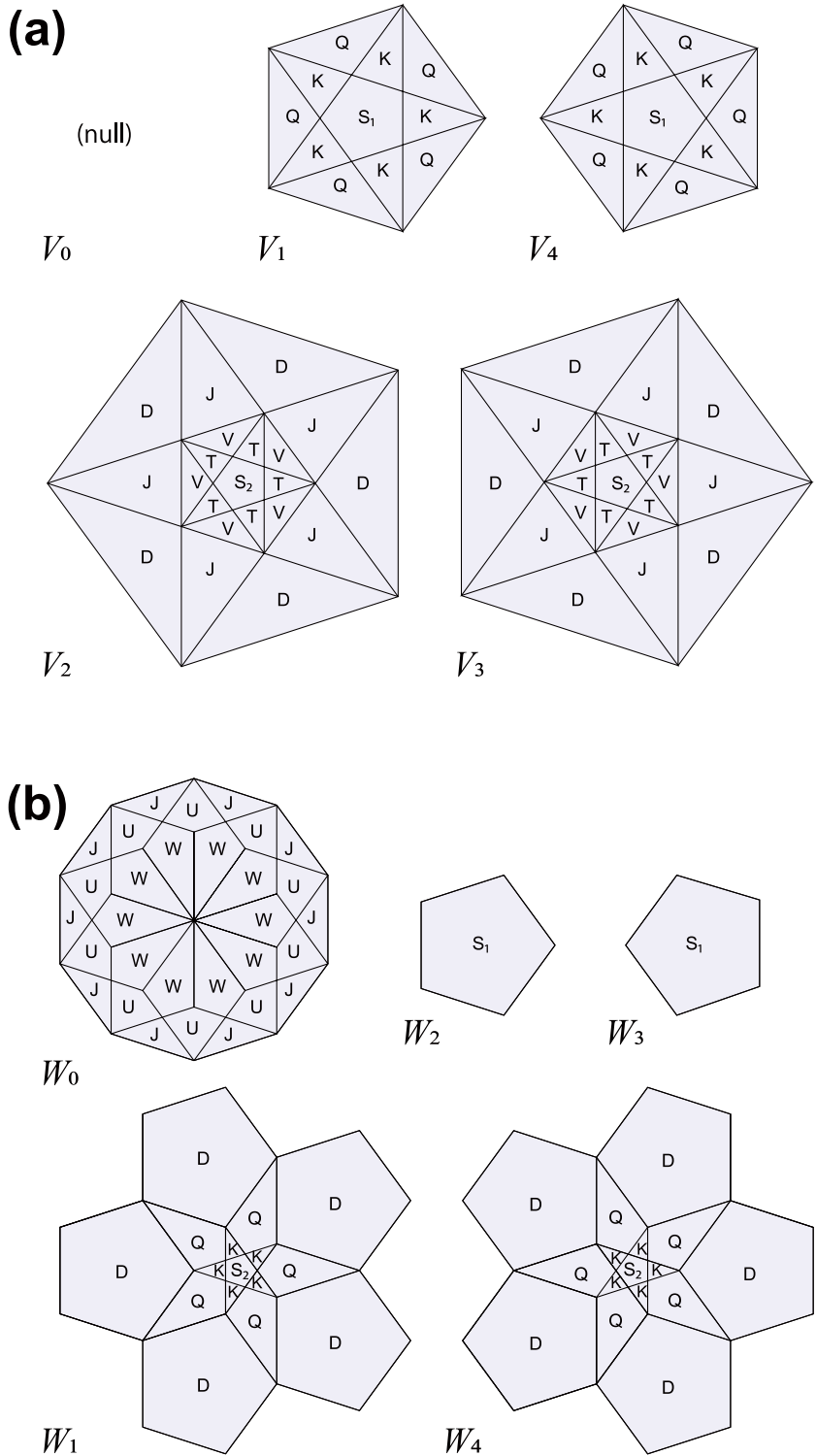

**Figure 3:** The ADs for the P3 tiling (a) and those for the P4 tiling (b). As indicated, each AD is segmented by solid lines into sections corresponding to the distinct vertex environments presented in *Figure 2*.

**(a)**

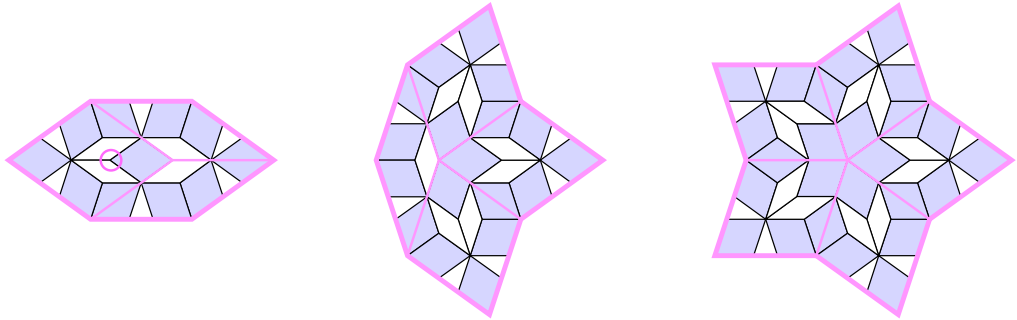

**(b)**

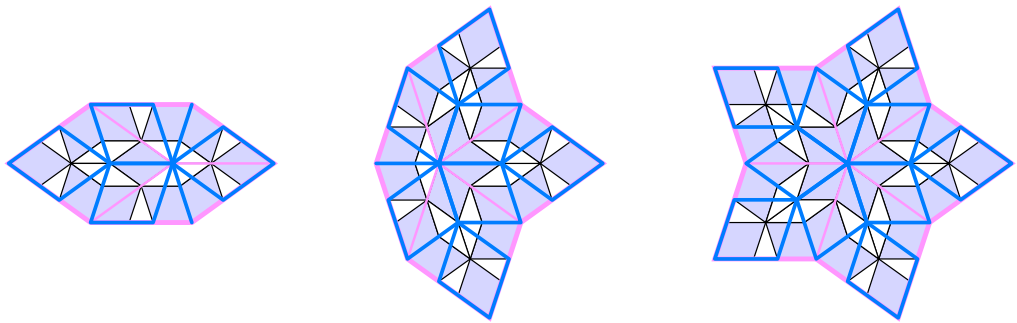

*Figure 4:* (a) Hexagon, boat and star polygons cut out from the P4 tiling in Figure 1 by edges of length $\tau^2$ connecting $S_1$-type vertices. Each polygon is decomposed into rhombuses (shown by pink solid lines) by choosing an appropriate internal vertex. The hexagon allows for two such choices, with another choice indicated by a pink circle. (b) The decomposition of the three polygons into smaller rhombuses with an edge length of $\tau$ (shown by blue solid lines) by deflating the pink rhombuses.

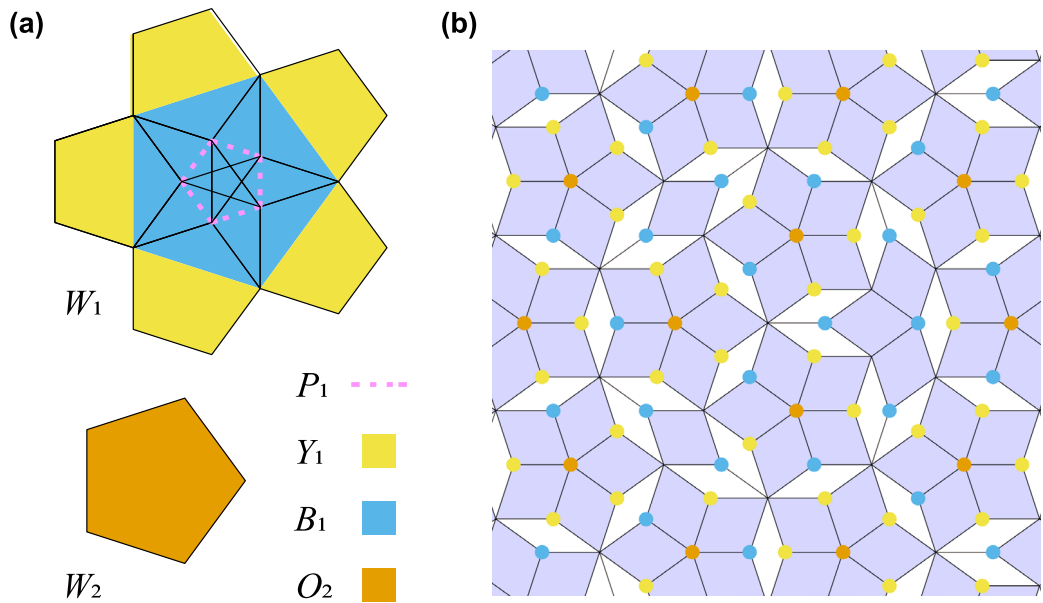

*Figure 5:* (a) Pentagonal sub-domains in $W_1$ and $W_2$. The sub-domain $P_1$, outlined with pink dashed lines, corresponds to the P3 tilings with edge length $\tau^2$, while the blue-painted sub-domain $B_1$ corresponds to the P3 tiling with edge length $\tau$. The yellow-painted sub-domain $Y_1$, consisting of five trapezoids, is the complement of $B_1$ in $W_1$. The corresponding images for $W_4$ and $W_3$ (not shown) are mirror images of $W_1$ and $W_2$, respectively. (b) The P4 tiling with vertices coloured as defined in (a). The blue and orange vertices correspond to the vertices of the $\tau$-scaled P3 tiling; see also *Figure 7*(a).

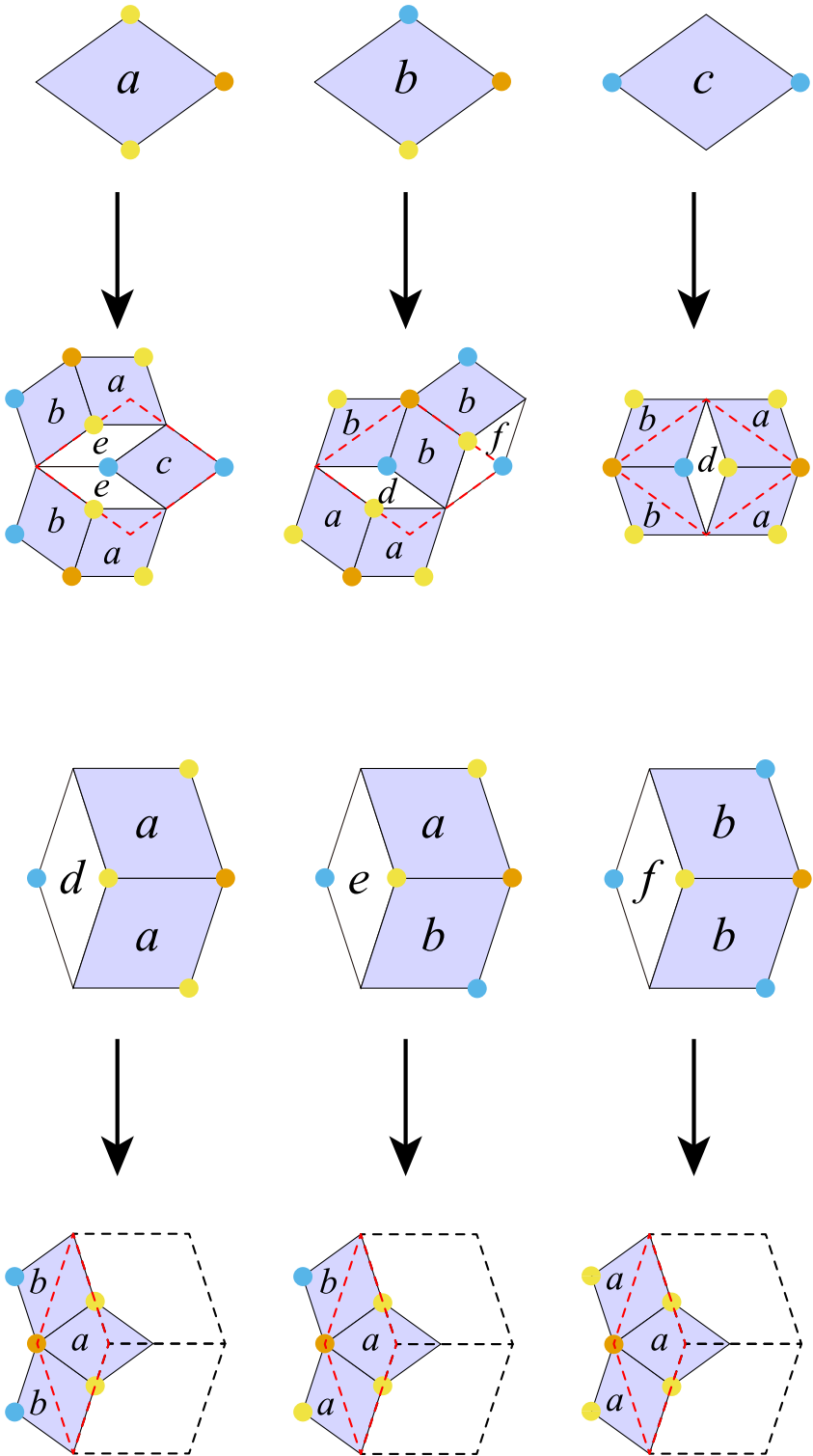

***Figure 6:*** **The classification of rhombuses in the P4 tiling and their deflation rules. The thick rhombuses are classified into three types (a, b, and c) based on the colouring of their vertices, whilst the thin rhombuses are also classified into three types (d, e, and f), differentiated by the types of two adjacent thick rhombuses that share a yellow vertex.**

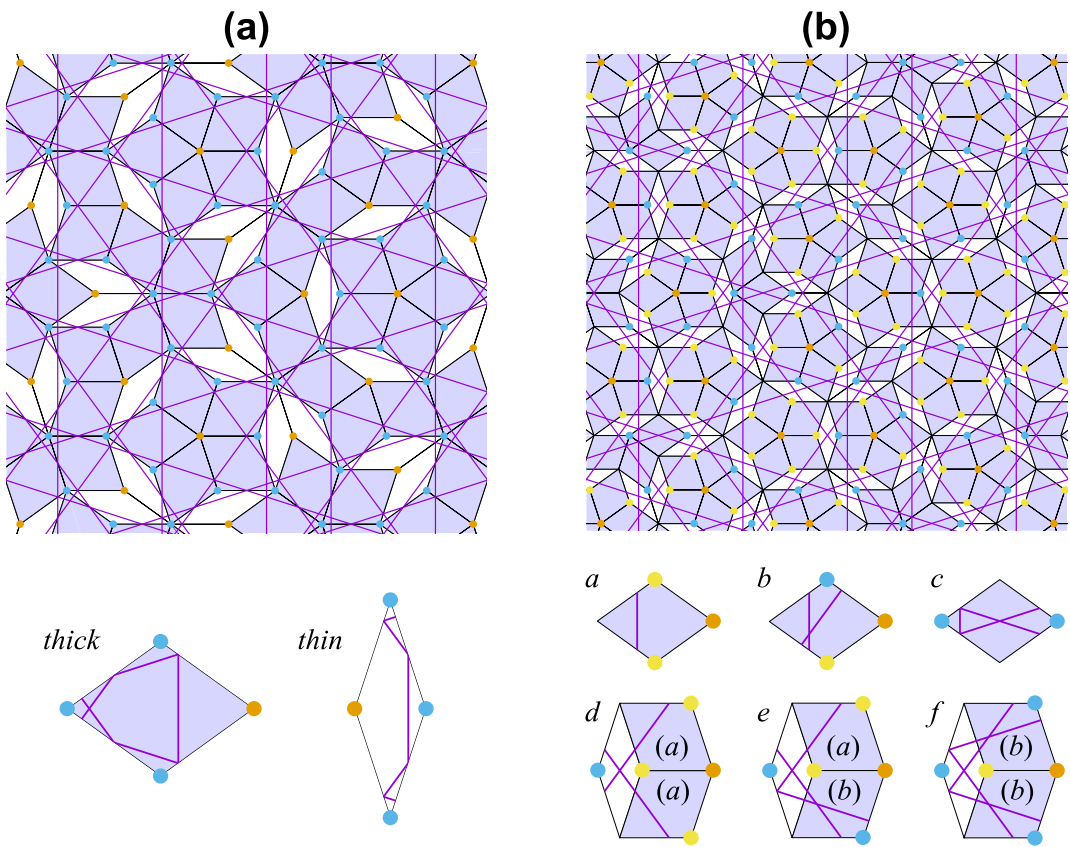

**(a)** **(b)**

*thick* *thin*

*a* *b* *c*

*d* (a) (a) *e* (a) (b) *f* (b) (b)

**Figure 7:** The Ammann bar grid (straight lines in purple) drawn over the $\tau$-scaled P3 tiling (a) and the P4 tiling (b). The prototiles of each tiling are shown at the bottom, decorated with the corresponding Ammann bars.

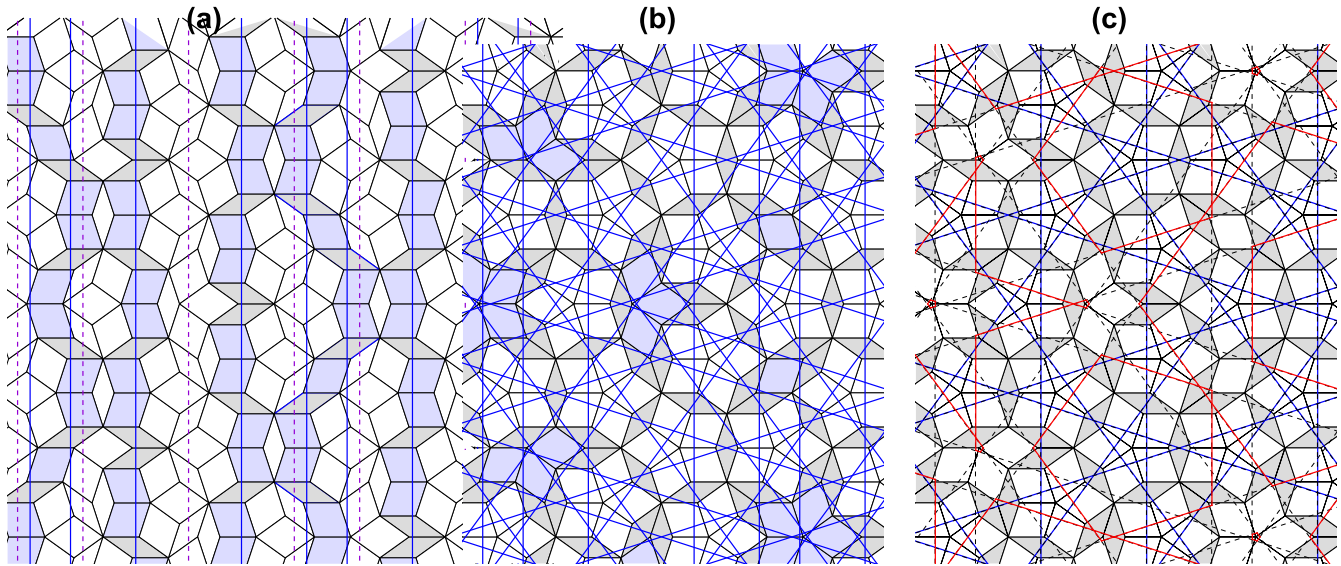

*Figure 8:* **The construction of the dual grid for the P4 tiling. (a) Vertical blue lines passing through the centers of thin rhombuses with horizontal edges, shaded in grey, are introduced as grid lines associated with the P4 tiling. The corresponding vertical arrays of rhombuses are highlighted (with thick rhombuses in blue). The dashed purple lines represent the Ammann lines in the same direction. (b) The five sets of parallel grid lines in all the five edge-normal directions are constructed in the same manner as in (a). The thin rhombuses are shaded in grey, while the thick rhombuses of type *c* are highlighted in blue; the latter kind of rhombuses occur exclusively at 72° intersections of $\ell_3$-pairs. (c) The dual grid for the P4 tiling is illustrated as consisting of straight blue lines and folded red lines, with dashed black lines indicating the inactive segments of the $\ell_3$-pairs, which are otherwise omitted.**

# Supplementary information for "A novel variant of rhombic Penrose tiling"

**Nobuhisa Fujita[1] and Komajiro Niizeki[2]**

[1] Institute of Multidisciplinary Research for Advanced Materials, Tohoku University, Sendai 980-8577, Japan
[2] Professor Emeritus, Department of Physics, Graduate School of Science, Tohoku University, Sendai 980-8578, Japan

E-mail: nobuhisa@tohoku.ac.jp

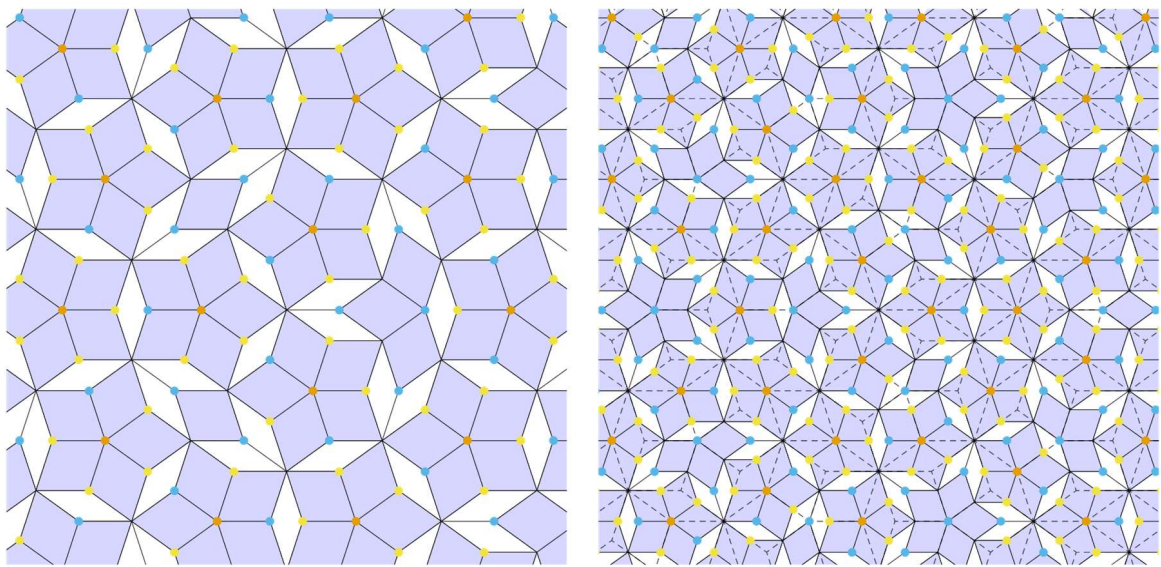

**Supplementary Figure 1:** Inflation/deflation symmetry of the P4 tiling. The $\tau$-scaled version of the P4 tiling (with edge length $\tau$), shown in the left panel, is transformed into the non-scaled version (with edge length 1) in the right panel, following the deflation rules illustrated in Figure 6 of the main text. In the right panel, dashed lines correspond to the rhombic edges shown in the left panel.

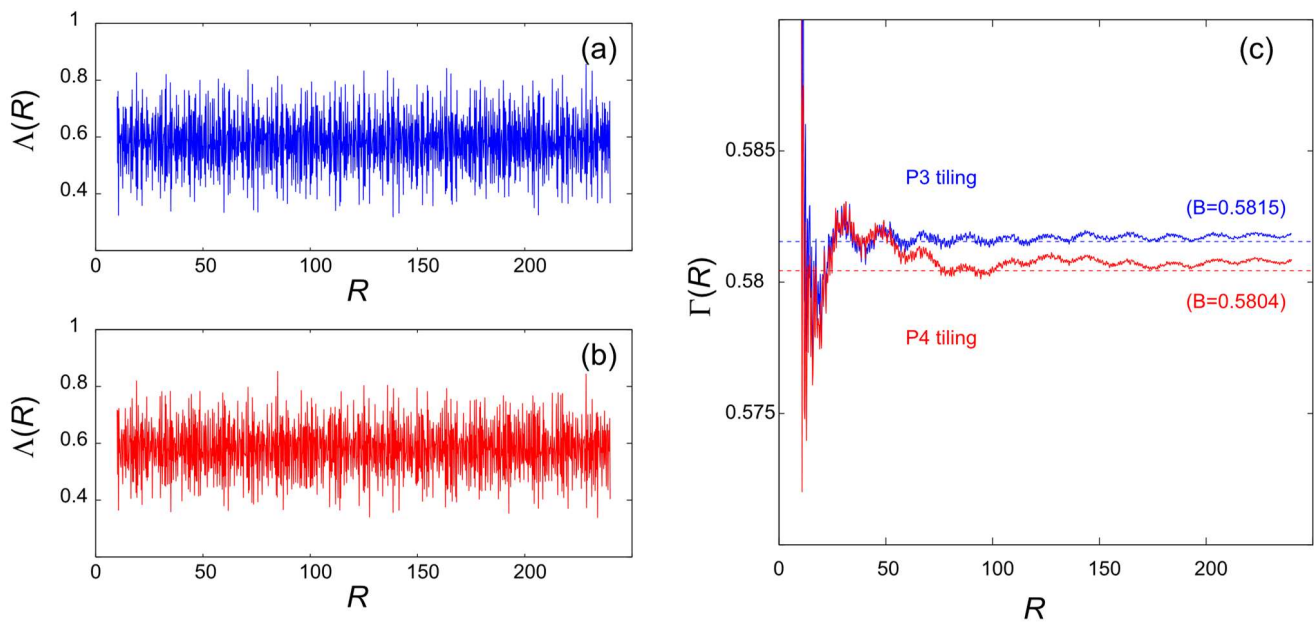

**Supplementary Figure 2: :** (a, b) The function $\Lambda(R) \coloneqq \sigma^2(R)/R$ evaluated for the P3 and P4 tilings, respectively. In both cases, the data exhibit bounded fluctuations around a constant value (i.e., the order metric $B$) as $R \to \infty$. (c) The running average of $\Lambda(R)$, defined as $\Gamma(R) \coloneqq \frac{1}{R-R_0} \int_{R_0}^{R} \Lambda(R')dR'$, is shown for the P3 (blue) and P4 (red) tilings, with $R_0 = 10$. For each tiling, an estimate of the order metric $B$ was obtained by fitting $\Lambda(R)$ within the range $10 < R < 240$ using the asymptotic functional form $\Lambda(R) \approx B + C/R$ using the linear least squares method. The corresponding order metrics $B$ are indicated by dashed lines in panel (c). The normalized order metrics $B/\sqrt{\phi}$, where $\phi = \pi\rho/4$ and $\rho$ is the number density of points, are 0.5914 and 0.5903 for the P3 and P4 tilings, respectively. The value obtained for the P3 tiling is consistent with previous results in the literature [1, 2].